\documentclass[aps,pra,10pt,twocolumn,nofootinbib,floatfix,superscriptaddress]{revtex4-2}

\usepackage[a4paper,left=1.8cm,right=1.8cm,top=2.5cm,bottom=3cm]{geometry}
\usepackage{graphicx}
\usepackage{amsmath}
\usepackage{amssymb}
\usepackage{mathtools}
\usepackage{enumitem}
\usepackage[dvipsnames]{xcolor} 
\definecolor{myblue}{RGB}{34,31,150} 
\usepackage[breaklinks=true,colorlinks=true,linkcolor=myblue,urlcolor=myblue,citecolor=myblue]{hyperref}

\begin{document}

\title{Adaptive, symmetry-informed Bayesian metrology for precise quantum technology measurements}

\author{Matt Overton}
\thanks{The first two authors contributed equally to this work. Corresponding author: \href{mailto:jesus@rubiojimenez.com}
{jesus@rubiojimenez.com}}
\affiliation{School of Physics and Astronomy, University of Nottingham, University Park, Nottingham NG7 2RD, United Kingdom}

\author{Jes\'{u}s Rubio} 
\thanks{The first two authors contributed equally to this work. Corresponding author: \href{mailto:jesus@rubiojimenez.com}
{jesus@rubiojimenez.com}}
\affiliation{School of Mathematics and Physics, University of Surrey, Guildford GU2 7XH, United Kingdom}

\author{Nathan Cooper}
\affiliation{School of Physics and Astronomy, University of Nottingham, University Park, Nottingham NG7 2RD, United Kingdom}

\author{Daniele Baldolini}
\affiliation{School of Physics and Astronomy, University of Nottingham, University Park, Nottingham NG7 2RD, United Kingdom}

\author{David Johnson}
\affiliation{School of Physics and Astronomy, University of Nottingham, University Park, Nottingham NG7 2RD, United Kingdom}

\author{Janet Anders}
\affiliation{Department of Physics and Astronomy, University of Exeter, Stocker Road, Exeter EX4 4QL, United Kingdom}
\affiliation{Institute of Physics and Astronomy, University of Potsdam,  14476 Potsdam, Germany}

\author{Lucia Hackerm\"{u}ller}
\affiliation{School of Physics and Astronomy, University of Nottingham, University Park, Nottingham NG7 2RD, United Kingdom}

% Abstract
\begin{abstract}
High precision measurements are essential to solve major scientific and technological challenges, from gravitational wave detection to healthcare diagnostics. Quantum sensing delivers greater precision, but an in-depth optimisation of measurement procedures has been overlooked. Here we present a systematic strategy for parameter estimation in the low-data limit that integrates experimental control parameters and natural symmetries. The method is guided by a Bayesian quantifier of precision gain, enabling adaptive optimisation tailored to the experiment. We provide general expressions for optimal estimators for any parameter. The strategy’s power is demonstrated in a quantum technology experiment, in which ultracold caesium atoms are confined in a micromachined hole in an optical fibre. We find a five-fold reduction in the fractional variance of the estimated parameter, compared to the standard measurement procedure. Equivalently, our strategy achieves a target precision with a third of the data points previously required. Such enhanced device performance and accelerated data collection will be essential for applications in quantum computing, communication, metrology, and the wider quantum technology sector.
\end{abstract}

\maketitle

From advances in gravitational wave detection \cite{Hogan2016,AION} to applications such as brain pattern imaging \cite{Zhang2020}, magnetic sensing \cite{Degen2017,Boto2018} or inertial navigation \cite{Narducci2022}, quantum technologies are revolutionising fundamental and applied research. 
Rapid estimation of platform parameters, as well as precise state preparation and control, underpins this success, enabling, e.g., efficient interferometric measurements \cite{Lee2022,Adams2023} and high-fidelity readout within a limited coherence time in quantum computing \cite{Graham2022,Wintersperger2023}.
Estimation efficiency \cite{Deng2024,Len2022} is critical for advancing quantum technologies \cite{horodecki2022five,Degen2017} and for testing the fundamental laws of nature \cite{turyshev2007, kaltenbaek2021quantum, belenchia2021quantum, ye2024essay, bass2024quantum}.

Local estimation theory is the standard for benchmarking measurement processes \cite{kay1993fundamentals, paris2009quantum, demkowicz2015quantum, montenegro2024review}. 
Here, the rate of information acquisition is increased by optimising the (quantum) Fisher information \cite{jaynes2003probability, liu2020quantum,len2022quantum, arvidssonshukur2020quantum}, and optimal sensitivity is achieved when the Cramér-Rao bound is saturated. 
This often requires an asymptotically large number of repetitions\,\cite{kay1993fundamentals, demkowicz2015quantum}, necessitating long acquisition times and possibly leading to information loss at the data processing stage \cite{rubio2018quantum, valeri2020experimental, morelli2021bayesian, rubio2021global, salmon2023only, meyer2023quantum}. 
In contrast, Bayesian estimation theory \cite{jaynes2003probability, linden2014bayesian} goes beyond by maximising information acquisition for {\it any sample size} \cite{rubio2021global}. Its effectiveness has been demonstrated within a limited number of atomic, molecular, and optical experiments, primarily for phase estimation \cite{brivio2010experimental, paesani2017experimental, lumino2018experimental, kaubruegger2021quantum, gianani2020assessing, glatthard2022optimal, gerster2022experimental, valeri2023experimental, oliveira2023fluctuation, cimini2024benchmarking, belliardo2024optimizing, hewitt2024controlling, ma2025adaptive}. 
Despite the clear advantages and well-established foundations of the Bayesian approach,  \cite{helstrom1976quantum, jaynes2003probability, agostini2003bayesian, 
linden2014bayesian, 
macieszczak2014bayesian,
demkowicz2015quantum, demkowicz2020multiparameter, bavaresco2024designing, rubio2024first}, experimental adoption for quantum sensing applications remains limited. 

\begin{figure}[t]
    \centering
    \includegraphics[trim={0cm 0cm 0cm 0cm},clip,width=0.9\linewidth]{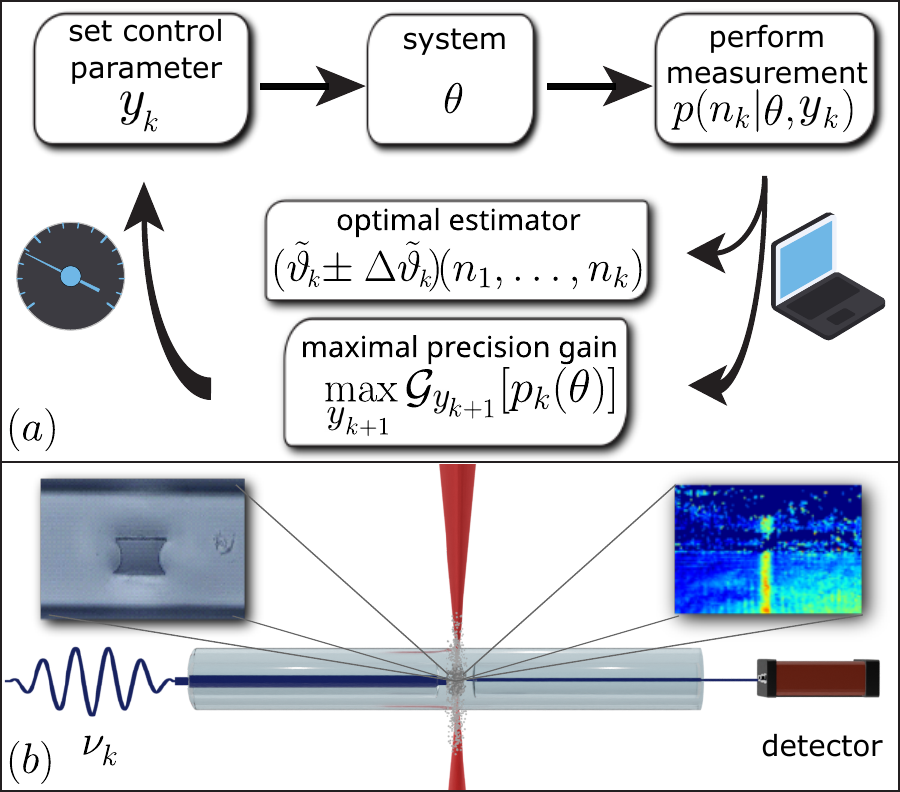}
    \caption{\sf 
    (a) General adaptive, symmetry-informed, Bayesian estimation strategy. 
    Information gained about system parameter $\theta$ from a detected signal $n_k$ determines a new optimal control parameter $y_k$ that maximises information gain in the next measurement shot, $k+1$. 
    (b) Application to an ensemble of cold Cs atoms confined in an optical dipole trap (red cones) within a microscopic hole (left inset) that intersects the core of an optical fibre.  
    Atom number is probed via photon-absorption detection of light travelling along the fibre.
The light's frequency $\nu_k$ is the control parameter to be optimised. 
    The final estimate of atom number is sequentially improved as photon-counts are recorded.
    The right inset shows an absorption image of the atoms passing through the optical fibre.  
    }      
    \label{fig:adaptive-scheme}
\end{figure}

Here we present a systematic strategy for experimental sensing accelerating information acquisition, or equivalently, enabling higher precision. 
This is achieved through Bayesian optimisation of experimental control parameters $y$ in order to estimate an unknown parameter $\Theta$, as illustrated in Fig.~\ref{fig:adaptive-scheme}(a).
Unlike standard adaptive strategies \cite{polino2020photonic, mehboudi2022fundamental, kurdzialek2023using, smith2024adaptive}, our approach is guided by a precision gain quantifier that incorporates the problem's natural symmetries, ensuring both consistency and maximal information extraction.
Our strategy applies to experimental parameters $\Theta$ that are not phases, such as atom number\,\cite{Su2025}, decay rates\,\cite{Zhang2025,rubio2023quantum}, or relative weights\,\cite{jaynes2003probability,rubio2023quantum}, and significantly expands beyond conventional phase estimation. This transferable strategy can be readily adopted to boost performance across the quantum technology sector, and will provide significant benefit when probing challenging regimes, such as in searches for new fundamental physics with wide \emph{a priori} bounds for unknown parameters \cite{AION, Schrinski2023, Clements2024}.

We further demonstrate this strategy's efficacy in an example experiment, where a cold atomic cloud is trapped inside an optical fibre \cite{daRos2020, Cooper2019}; see Fig.~\ref{fig:adaptive-scheme}(b), and atom number is the system parameter to be estimated. 
Controlling the probing laser frequency according to our adaptive symmetry-informed Bayesian protocol significantly enhances information acquisition speed and precision.

\emph{Adaptive symmetry-informed estimation}.---Efficient estimation amounts to solving an optimisation problem \cite{helstrom1976quantum, jaynes2003probability, kurdzialek2023using, maclellan2024end}. 
We present a general adaptive symmetry-informed Bayesian strategy for parameter estimation in experiments where measurement precision depends on one or more control parameters and estimated parameters are not necessarily phases. 

Let $\theta \in [\theta_{\mathrm{min}},\theta_{\mathrm{max}}]$ be a {\it hypothesis} for an unknown parameter $\Theta$ to be determined indirectly through a measurement with outcome $n$. 
This outcome is related with $\theta$ and possibly a control parameter $y$ through a likelihood function
$p(n|\theta, y)$.
For $k$ outcomes $\boldsymbol{n} = (n_1, \dots, n_k)$ measured using control parameters $\boldsymbol{y} = (y_1, \dots, y_k)$, 
the likelihood is $p(\boldsymbol{n}|\theta, \boldsymbol{y}) = \prod_{i=1}^k p(n_i|\theta, y_i)$.
Based on this, the default choice for an estimator for $\Theta$ is the maximum likelihood estimator (MLE) $\tilde{\theta}_{\boldsymbol{y}}(\boldsymbol{n}) = \mbox{argmax}_{\theta}  [p(\boldsymbol{n}|\theta, \boldsymbol{y})]$. However, the MLE is known to be efficient only in the regime of large sample sizes \cite{demkowicz2015quantum, kay1993fundamentals}.

This motivates adopting a Bayesian approach. 
Here, a prior probability $p(\theta)$, which encodes the initial information, is \emph{updated} via Bayes's theorem as 
\begin{equation}
   p(\theta) \mapsto p(\theta|\boldsymbol{n}, \boldsymbol{y}) \propto p(\theta) p(\boldsymbol{n}|\theta, \boldsymbol{y})
\end{equation}
upon recording $\boldsymbol{n}$, where
\begin{equation}
    p(\theta|\boldsymbol{n}, \boldsymbol{y}) \propto p(\theta) \prod_{i=1}^k p(n_i|\theta, y_i)
    \label{eq:posterior}
\end{equation}
is the posterior probability. 
This probability contains all the information available about $\Theta$ \cite{jaynes2003probability, glatthard2022optimal}. 

In addition to using Bayes's theorem, one must respect the symmetry of the problem. 
For example, while phases obey circular symmetry, $\theta' = \theta + 2\pi$ \cite{demkowicz2011optimal}, temperature usually appears in ratios of energies, which are invariant under transformations 
$\theta' = \gamma \theta$, for positive $\gamma$ \cite{rubio2021global}.
A symmetry function $f$ that weights the posterior probabilities \eqref{eq:posterior} can incorporate such symmetries into the measurement strategy, thereby enforcing the desired symmetry in the optimal estimation.
For a given $f$, measurement outcomes $\boldsymbol{n}$, and control parameters $\boldsymbol{y}$, the symmetry-informed estimator for $\Theta$ is
\begin{equation}
  \tilde{\vartheta}_{\boldsymbol{y},f}(\boldsymbol{n})  = f^{-1} \left[\int d\theta\,p(\theta|\boldsymbol{n}, \boldsymbol{y}) f(\theta)\right] . \label{eq:opt-est} 
\end{equation}
This estimator is \emph{optimal} under quadratic losses of the form $[f(\tilde{\theta}) - f(\theta)]^2$  \cite{rubio2024first}.
When $f(\theta) = \theta$, the estimator is simply the mean of the posterior probability, suitable for location parameters with translation symmetry $\theta' = \theta + \gamma$, with arbitrary $\gamma$. Other symmetries necessitate different estimator functions. 
Accounting for the symmetries of individual applications is thus an essential step in generalising this estimation strategy.
Furthermore, the empirical error associated with this estimator is
\begin{equation}
  \Delta \tilde{\vartheta}_{\boldsymbol{y},f}(\boldsymbol{n}) = \frac{\sqrt{\mathcal{L}_{\boldsymbol{y},f} (\boldsymbol{n})}}{|f'[\tilde{\vartheta}_{\boldsymbol{y},f}(\boldsymbol{n})]|},
  \label{eq:opt-err}
\end{equation}
where  
\begin{equation}
     \mathcal{L}_{\boldsymbol{y},f}({\boldsymbol{n}}) = \int d\theta\, p(\theta|{\boldsymbol{n}}, \boldsymbol{y}) f(\theta)^2  
     - f[\tilde{\vartheta}_{\boldsymbol{y},f}(\boldsymbol{n})]^2
     \label{eq:outcome-loss}
\end{equation}
is the loss incurred by processing $(\boldsymbol{n}, \boldsymbol{y})$ via Eq.~\eqref{eq:opt-est}.

Eqs.~\eqref{eq:opt-est} and ~\eqref{eq:opt-err} provide maximum information for any parameter range or sample size and enable experimental symmetry-informed estimation through appropriate choice of $f$. For example, the likelihood may be invariant under a symmetry group \cite{kass1996the}, or we may enforce invariance under all reparametrisations using information geometry \cite{linden2014bayesian, amari2016}. 
These symmetries can be shown to determine maximum ignorance priors of the form \cite{rubio2024first}
\begin{equation}
    p_{\mathrm{MI}}(\theta) \propto f'(\theta),
    \label{eq:mi-prior}
\end{equation}
where $p_{\mathrm{MI}}(\theta)$ follows from the invariance condition \cite{linden2014bayesian}
\begin{equation}
    p_{\mathrm{MI}}(\theta')\, d\theta' = p_{\mathrm{MI}}(\theta)\, d\theta
    \label{eq:invariant-condition}
\end{equation}
with $\theta'$ and $\theta$ related by an arbitrary symmetry.
The function $f$ entering Eqs.~\eqref{eq:opt-est} and \eqref{eq:opt-err} is obtained from Eqs.~\eqref{eq:mi-prior} and \eqref{eq:invariant-condition}.
This rationale is applied later, in the derivation of Eqs.~\eqref{eq:prior-mi} and \eqref{eq:log-function}.

The information loss \eqref{eq:outcome-loss} can now be {\it adaptively} reduced by selecting $y_k$ for the $k$-th measurement to maximise the precision gain \cite{rubio2024first, glatthard2022optimal}
\begin{equation}
    \mathcal{G}_{y_k, f} = \sum_n p(n|y_k)\,f[\tilde{\vartheta}_{y_k,f} (n)]^2.
    \label{eq:precision-gain}
\end{equation}
Here, $p(n|y_k) = \int d{\theta}\,p_{k-1}(\theta) p(n|\theta, y_k)$, with
\begin{equation}
    p_{k-1}(\theta) \coloneqq
\begin{cases}
    p(\theta) & \text{if } k = 1 \\
    p(\theta) \prod_{i = 1}^{k-1} p(n_i|\theta, y_i) & \text{for } k > 1,
\end{cases}
\label{eq:adaptive-general}
\end{equation}
and $\tilde{\vartheta}_{y_k,f} (n)$ is defined as in Eq.~\eqref{eq:opt-est} using $p(\theta|n, y_k) \propto p_{k-1}(\theta) p(n|\theta, y_k)$.
By construction, $\mathcal{G}_{y_k, f}$ guarantees adaptive optimisation that respects the symmetry of the problem through $f$.
Eqs.~\eqref{eq:opt-est}, \eqref{eq:opt-err}, and \eqref{eq:precision-gain} constitute our first result and are derived in the Supplemental Material.

\emph{Application to experimental atom number metrology}.---
We now apply the equations derived above to a specific experiment to demonstrate their power.
The example is a cold atom setup~\cite{daRos2020}, where Cs atoms are trapped within a hole cut into an optical fibre, see Fig.~\ref{fig:adaptive-scheme}(b). 
This setup is a promising platform \cite{daRos2020, Cooper2019} for the integration of cold-atom based sensors, quantum memories and multi-qubit gates within existing photonic circuit architectures \cite{Bogaerts2020}.
The task is to determine the number of atoms $N$ trapped within the fibre. Previously this required $30-100$ shots per estimate and was the main rate-limiting step.
Requiring many repetitions, or long measurement times, is typical, e.g., for atom-based systems such as atom-nanofibre experiments\,\cite{nanofiber,Corzo2019}, precision magnetometers\,\cite{Troullinou2023} and atom interferometry\,\cite{Zhou2024}. 
Here we address this problem with an adaptive strategy based on Eq.~\eqref{eq:precision-gain} and an optimal atom number estimator $\widetilde{N}_{\boldsymbol{\nu}}(\boldsymbol{n}) \pm \Delta \widetilde{N}_{\boldsymbol{\nu}}(\boldsymbol{n})$ built upon Eqs.~\eqref{eq:opt-est} and ~\eqref{eq:opt-err}. 

Near-resonant light propagating through the fibre is absorbed by the atoms in the intersection at a rate proportional to \cite{daRos2020}
\begin{equation}
  \zeta_{\nu} = \frac{\Gamma^2}{\Gamma^2+4(\nu-\nu_{\mathrm{r}})^2},
  \label{zeta}
\end{equation}
where $\nu$ is the laser frequency, and  $\nu_{\mathrm{r}}$ and $\Gamma$ are the resonant frequency and natural linewidth of the atomic transition, respectively.
The light intensity is kept low compared to the atomic saturation intensity, leading to an expected photon count of
\begin{equation}
    \bar{n}_{\nu}(\Phi, \Theta) = \Phi\,\mathrm{e}^{- \zeta_{\nu} \Theta}.
    \label{eq:av-count}
\end{equation}
Here, $\Phi$ is the expected photon number with no atoms present, $\mathrm{exp}(- \zeta_{\nu}\Theta)$ is the transmittance of the medium, and $\Theta$ its on-resonance optical depth. 
Both $\Phi$ and $\Theta$ are initially unknown. 
The atom number $N$ we seek to determine is proportional to the optical depth $\Theta$, i.e. $N = \kappa\,\Theta$ \cite{daRos2020}, with constant $\kappa$ determined by the mode area of the waveguide and the on-resonance atomic absorption cross-section. 
The standard method for estimating $N$, employed in most setups using optical absorption to detect or image cold atoms \cite{Raab1987, Turner_2004, ketterle1999bec, Ness2020, Vibel_2024}, is to count photons with and without atoms loaded into the hole ($n = n_{\mathrm{a}}$ and $n = n_{\mathrm{b}}$, respectively), and use the MLE $\tilde{\theta}_{\nu}(\boldsymbol{n}) = {\kappa \over \zeta_\nu} \log {\langle n_{\mathrm{a}} \rangle \over \langle n_{\mathrm{b}} \rangle}$, with error propagated from this formula.
Here, $\nu$ is set on resonance with $\nu_r$ (i.e., $\zeta_\nu = 1$).
This estimator results from the likelihood associated with photon count statistics,
\begin{equation}
  p(n \mid \Phi, \Theta, \nu) = P(n \mid \bar{n}_{\nu}(\Phi, \Theta)),
  \label{eq:prob-atoms}
\end{equation}
where $P(n \mid z) = z^n \mathrm{e}^{-z}/n!$, which arises because ordinary (unsqueezed) light exhibits a Poissonian distribution.

Instead, we will now apply the new protocol with the laser frequency $\nu$ as the control parameter $y$ in Eq.~\eqref{eq:opt-est}. 
Symmetry-informed estimation is applied directly to the optical depth $\Theta$, with $\Phi$, the photon-number without atoms, featuring as an undetermined nuisance parameter.

\begin{figure*}[t]
\centering
\includegraphics[trim={0cm 0cm 0cm 0cm},clip,width=\linewidth]{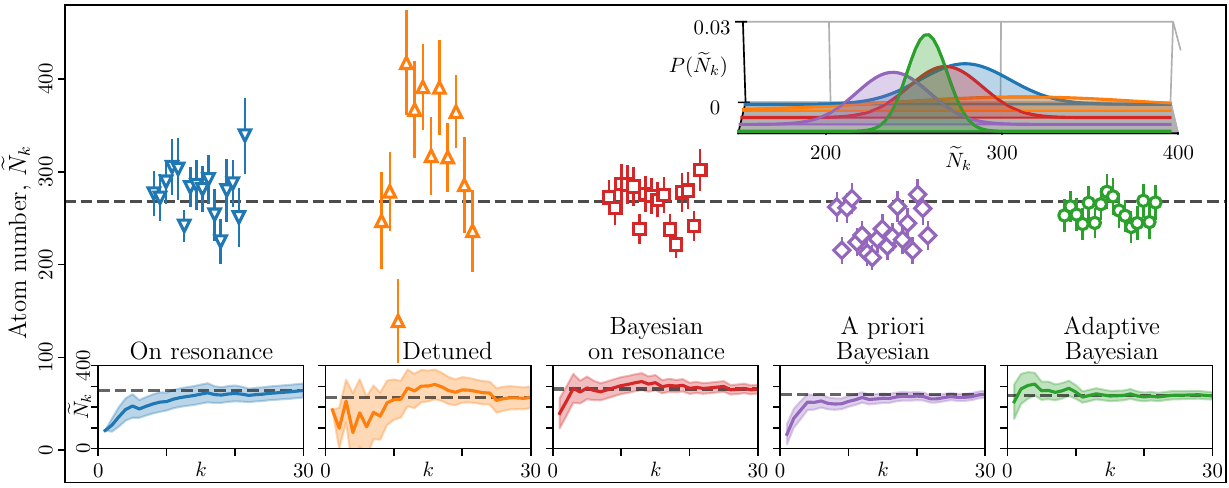}
\caption{ \sf
Atom number estimates $\tilde{N}_k$ resulting from measurements using the standard (blue and orange) and Bayesian estimation strategies (red, purple and green) described in the text. 
Each estimate employs $k=30$ measurement shots. 
Error bars are given by the standard error for each $30$-shot measurement in the on-resonance and detuned cases and by the variance of Eq.~\eqref{eq:marginal-posterior} otherwise.
The lower insets show the progression of the estimate with increasing $k$, for a single measurement run, in each case. 
The shading denotes the aforementioned errors. 
The small differences in the mean value obtained with each method are consistent with normal experimental drifts in loading efficiency.
The dashed lines indicate final estimates in the insets and overall average across all methods in the main figure.
The inset in the top right corner gives a visual Gaussian representation, $P(\tilde{N}_k)$, of the mean and the standard deviation of the data presented in the main body of the figure for each method.
} 
\label{fig:res1}
\end{figure*}

Let $\theta \in [\theta_{\mathrm{min}},\theta_{\mathrm{max}}]$ and $\varphi \in [\varphi_{\mathrm{min}},\varphi_{\mathrm{max}}]$ be the hypotheses for the real $\Theta$ and $\Phi$, respectively.
When two observers run the experiment at frequency $\nu$, their average photon counts are comparable, but their initial hypotheses can differ due to prior ignorance.
That is, $\bar{n}_{\nu}(\varphi', \theta') = \bar{n}_{\nu}(\varphi, \theta)$, so that $\varphi'/\varphi = \exp[\zeta_{\nu}(\theta'-\theta)]$.
Equating both sides to a positive $\gamma$ reveals transformations 
\begin{equation}
    \varphi' = \gamma \varphi,\text{ } \quad
    \zeta_\nu \theta' = \zeta_\nu \theta + \log(\gamma)
\label{eq:symmetry}
 \end{equation}
under which Eq.~\eqref{eq:prob-atoms} is invariant.
These imply that the expected number of photons is a scale parameter, while the optical depth is a location parameter \cite{jaynes2003probability}.
Then, if both observers are equally ignorant, and given that $\nu$ does not inform $\Phi$ and $\Theta$, their priors must respect the aforementioned invariance as $p_{\mathrm{MI}}(\varphi', \theta') \, d\varphi' \, d\theta' = p_{\mathrm{MI}}(\varphi, \theta) \, d\varphi \, d\theta$.
As shown in the Supplemental Material, this leads to the multiparameter prior
\begin{equation}
    p_{\mathrm{MI}}(\varphi,\theta) = \left[ (\theta_{\mathrm{max}}-\theta_{\mathrm{min}}) \log\left(\frac{\varphi_{\mathrm{max}}}{\varphi_{\mathrm{min}}}\right) \varphi \right]^{-1}.
    \label{eq:prior-mi}
\end{equation}
Despite having no \emph{a priori} knowledge about the unknown parameters beyond the initial ranges, this prior is \emph{not} flat in $\varphi$. 
This is a consequence of the scale symmetry in Eq.~\eqref{eq:symmetry} \cite{kass1996the, jaynes2003probability}.

Given that Eq.~\eqref{eq:prior-mi} is a two-parameter prior, it yields two different symmetry functions. 
Marginalising over $\theta$ or $\varphi$ to obtain $f(\varphi)$ or  $f(\theta)$ from Eq.~\eqref{eq:mi-prior}, respectively, one finds 
\begin{equation}
    f(\varphi) = c_1 \log\left(\frac{\varphi}{c_2}\right) ~~~\mathrm{and} ~~~f(\theta) = c'_1\theta + c'_2,
    \label{eq:log-function}
\end{equation}
where $c_1$, $c_2$, $c'_1$ and $c'_2$ are free constants.
Since our focus is $\theta$, and $\varphi$ is a nuisance parameter, only the linear function 
is required. Note that in other contexts, such as quantum thermometry, the logarithmic function has been important \cite{rubio2021global, mehboudi2022fundamental, glatthard2022optimal}. 
In general, the function $f$ can take many forms; for example, this methodology could be applied equally effectively to experiments measuring a conversion efficiency such as Ref.~\cite{Naniyil_2022}, which is a weight parameter and would thus require $f(z) = 2~\mathrm{arctanh}(2 z-1)$  \cite{rubio2024first}.

Applying the symmetry function $f(\theta)$ from Eq.~\eqref{eq:log-function} and with $N \propto \Theta$, Eqs.~\eqref{eq:opt-est} and \eqref{eq:opt-err} deliver the optimal atom number estimator and its uncertainty $\widetilde{N}_{\boldsymbol{\nu}}(\boldsymbol{n}) \pm \Delta \widetilde{N}_{\boldsymbol{\nu}}(\boldsymbol{n})$ as the mean and variance of the marginalised posterior $p(\theta|\boldsymbol{n}, \boldsymbol{\nu})$, which is given by
\begin{equation}
    p(\theta|\boldsymbol{n}, \boldsymbol{\nu}) \propto \int_{\varphi_{\mathrm{min}}}^{\varphi_{\mathrm{max}}} 
    \frac{d\varphi}{\varphi}
    \prod_{i=1}^k p(n_i|\varphi,\theta,\nu_i),
    \label{eq:marginal-posterior}
\end{equation}
where $c'_1 = 1$ and $c'_2 = 0$ were chosen.
Eq.~\eqref{eq:marginal-posterior} respects the location symmetry associated with the optical depth $\theta$ and the scale symmetry associated with 
$\varphi$ through the ignorance prior \eqref{eq:mi-prior}.
This is our second result.

\emph{Absorption measurements at atom--optical interface}.--- 
To employ the protocol experimentally,
we measure the outcomes $\boldsymbol{n} = (n_{\mathrm{a}, 1}, n_{\mathrm{b}, 1}, \dots, n_{\mathrm{a}, k}, n_{\mathrm{b}, k})$, with $1 \leq k \leq 30$, alternating photon counts with and without atoms and setting $p(n_i|\varphi, 0, \nu_i)$ in Eq.~\eqref{eq:marginal-posterior} for the latter.
Laser frequency is set to maximise the precision gain \eqref{eq:precision-gain} with $\tilde{\vartheta}_{\nu_k}(n) \propto \widetilde{N}_{\nu_k}(n)$, thus finding the frequencies $\boldsymbol{\nu} = (\nu_{1 \mathrm{a}}, \nu_{1 \mathrm{b}}, \dots, \nu_{k \mathrm{a}}, \nu_{k \mathrm{b}})$; see Fig.~S1 in the Supplemental Material.
We choose prior ranges $5 \leq \varphi \leq 20$ and $0 \leq \theta \leq 8$.
For the Cs D$_2$ $F=4 \rightarrow F'=5$ transition, the physical values are $\nu_\mathrm{r} = 351.721 961$\,THz, $\Gamma = 2\pi\cdot5.234$\,MHz, and $\kappa = 84.9$~\cite{steck2024CsDLine, daRos2020}.
All experimental parameters other than $\nu$ are kept constant.
Eqs.~\eqref{eq:precision-gain} and \eqref{eq:marginal-posterior} are rapidly evaluated between shots by caching function results, managing tolerances, and adaptively truncating probabilities.

The effectiveness of the adaptive, symmetry-informed Bayesian strategy is quantified in comparison to four alternative methods: 
The first ("on resonance'') is the standard method \cite{daRos2020} discussed above.
The second method (``detuned'') is analogous, but with the laser frequency detuned below the atomic resonance, to $\nu_\mathrm{r}-\nu=5$\,MHz.
The third method (``on resonance Bayesian'') uses resonant light, $\nu=\nu_\mathrm{r}$, but adopts the Bayesian estimator from Eq.~\eqref{eq:marginal-posterior}. 
In the fourth method (``{\it a priori} Bayesian''), the laser frequency is chosen to maximise the precision gain \eqref{eq:precision-gain} for $k = 1$. This \emph{a priori} optimisation involves no outcomes \cite{rubio2018quantum, glatthard2022optimal}, and here results in a detuning of $\nu_\mathrm{r}-\nu=1.95$\,MHz.

Fig.~\ref{fig:res1} shows atom number estimates yielded by the on-resonance (blue inverted triangles), detuned (orange triangles), on-resonance Bayesian (red squares), \emph{a priori} Bayesian (purple rhombuses), and adaptive Bayesian (green circles) methods.
We repeat each method  $m$ times and quantify its performance by the empirical noise-to-signal ratio (NSR) 
$\mathrm{Var} (\widetilde{N}_k)/\langle \widetilde{N}_k\rangle^2$ \cite{glatthard2022optimal}, where $\langle \widetilde{N}_k\rangle = \frac{1}{m} \sum_{j=1}^{m} \widetilde{N}_{k,j}$ and $\widetilde{N}_{k,j}$ denotes the $k$-th estimate during the $j$-th repetition.
This figure of merit is appropriate as it is agnostic to the specific methodology employed.
The results, given in Tab.~\ref{tab:result_summary}, clearly show marked differences in performance between the strategies.
While the two non-adaptive Bayesian protocols and the on-resonance method yield comparable NSRs of just under $1\%$, and all three outperform the detuned method by an order of magnitude, the adaptive protocol surpasses them all with a NSR of $<0.2\%$.
Notably, the latter outperforms the on-resonance method---which is the current standard for cold atom platforms---by a factor of $4.68$.
This approximately five-fold enhancement constitutes our third result.

\begin{table}[t]
\centering
\begin{tabular}{l c c c c c }
\hline
\hline
& On & & On reso. &  \emph{A priori}  &   {\bf Adaptive} \\[-0.5ex]
&  resonance & Detuned &  Bayesian & Bayesian &  {\bf Bayesian} \\
\hline
NSR in $\%$  &  $0.89$ &  $6.10$ & $0.62$ & $0.77$ & {\bf 0.19} \\
$k_{\mathrm{min}}$    &  $13$  & $17$ & $12$ & $12$ & $\boldsymbol{8}$ \\
\hline  
\hline     
    \end{tabular}
    \caption{\sf Comparison of atom number estimation strategies in terms of noise-to-signal ratio for $k = 30$ measurement shots, as well as mean minimal number of shots $k_{\mathrm{min}}$ after which the estimate stays within 10\% of the final estimate, from the data shown in Fig.~\ref{fig:res1}.}
    \label{tab:result_summary}
\end{table}

\begin{figure}[t]
\includegraphics[trim={0cm 0cm 0cm 0cm},clip,width=\linewidth]{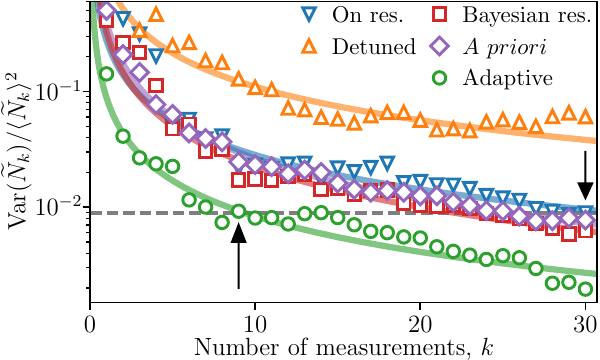}
    \caption{\sf Noise-to-signal ratio (NSR) versus shot number $k$ for the on-resonance (blue inverted triangles), detuned (orange triangles), on-resonance Bayesian (red squares), \emph{a priori} Bayesian (purple rhombuses), and adaptive Bayesian (green circles) strategies. The data points are obtained empirically from the distribution of $m$ estimates after each shot number $k$. 
    The dashed line indicates the NSR for on-resonance method with $30$ shots (downward facing arrow), which is achieved by our adaptive protocol with only $9$ shots (upward facing arrow).}
\label{fig:res2}
\end{figure}

Fig.~\ref{fig:res2} shows that these conclusions hold even for a low shot number. There, the NSR is represented as a function of $k$. A crucial implication is that the NSR achieved by the standard technique (blue inverted triangles) using $k=30$ shots is achieved by the adaptive protocol (green circles) with only $k=9$ shots, i.e., about one third of the resources.
This is our fourth result. 

The adaptive protocol was also found to converge faster.
When recording the mean number of measurement steps $k_{\mathrm{min}}$ needed before the current estimate settles to within $\pm10\,\%$ of the final estimate based on $30$ shots, we obtain the results listed in the last row of Tab.~\ref{tab:result_summary}.
The adaptive protocol achieves the specified fractional precision with approximately $40\%$ fewer shots (8 instead of 13) than the standard on-resonance method.

\medskip

\emph{Discussion}.-- We have developed a general symmetry-informed estimation strategy that will enable accelerated data collection across the quantum technology sector.
We have demonstrated its application to a test platform: atom number estimation in a fibre experiment.
The optimised protocol yielded a substantial enhancement in estimation precision (five-fold increase), or alternatively, increased acquisition speed (a third of data points), compared to current standards.
These gains make our approach a powerful tool for quantum sensing, metrology, and computation in data-limited regimes.
We have further shown that real-time adaptive optimisation is feasible with standard computational resources, even in multiparameter setups \cite{valeri2020experimental, glatthard2022optimal}.
In general, use of Eqs.~\eqref{eq:opt-est}, \eqref{eq:opt-err}, and \eqref{eq:precision-gain} expands beyond the application range of Bayesian phase estimation \cite{demkowicz2011optimal} to optimal Bayesian estimation of parameters with non-cyclical symmetry, such as temperature, atom number, and state occupation.
Given these generality and performance advantages, the presented strategy should be adopted in experiments ranging from precision spectroscopy and magnetometry to quantum state preparation and fundamental physics tests.

\medskip

\emph{Acknowledgements}.---JR gratefully thanks J. Boeyens, J. Glatthard, G. Barontini, V. Guarrera, M. Tsang and L. Correa for helpful discussions. 
JR acknowledges financial support from the Surrey Future Fellowship Programme.
JA gratefully acknowledges funding from EPSRC (EP/R045577/1,EP/T002875/1) and thanks the Royal Society for support.
The experiment was supported by the grant 62420 from the John Templeton Foundation, the IUK project No.133086, EPSRC grants EP/T001046/1, EP/R024111/1, EP/M013294/1, EP/Y005139/1 and  EP/Z533166/1, and by the European Commission grant ErBeStA (no.800942). 

\emph{Data and materials availability}.---All data upon which our conclusions are based are presented herein. All additional data related to the experimental system are available from the authors upon reasonable request.

\emph{For the purpose of open access, the authors have applied a `Creative Commons Attribution' (CC BY) licence to any Author Accepted Manuscript version arising from this submission.}

% References
\bibliography{refs}

% Supplement
\onecolumngrid
\setcounter{equation}{0}
\renewcommand\theequation{S\arabic{equation}}
\setcounter{figure}{0}
\renewcommand{\thefigure}{S\arabic{figure}}
\setcounter{table}{0} 
\renewcommand{\thetable}{S\arabic{table}}

\bigskip
\begin{center}
    \bf \large Supplemental material
\end{center}

Here we present the derivations underlying the theory of symmetry-informed estimation and details of the experimental implementation.
We also provide example data that help to illustrate the typical conditions under which the experiment operates and a discussion of efficient numerical evaluation of the optimal probe laser frequency. An approach to determining the likely utility of an adaptive Bayesian measurement strategy is also outlined, so that potential performance benefits can be assessed prior to investing resources into implementation. 
Finally, a discussion of pathways for future research that build upon the work in the main manuscript is given. 

\bigskip
\twocolumngrid
\noindent \textbf{Bayesian inference with incorporated symmetries}
\smallskip

A flexible approach to designing estimation strategies is minimising an information loss quantifier \cite{jaynes2003probability, helstrom1976quantum}. 
Let $\mathcal{L}(\tilde{\theta},\theta)$ quantify the loss incurred by an estimate $\tilde{\theta}$ should the hypothesis $\theta$ be correct. 
The mean loss is defined as
\begin{equation}
    \bar{\mathcal{L}}_{\boldsymbol{y}} = \int d\boldsymbol{n}\, p({\boldsymbol{n}}|\boldsymbol{y})  \mathcal{L}_{\boldsymbol{y}}({\boldsymbol{n}}),
    \label{eq:loss-fom}
\end{equation}
where $p({\boldsymbol{n}}|\boldsymbol{y}) = \int d{\theta}\,p(\theta)p({\boldsymbol{n}}|\theta, \boldsymbol{y})$ is the evidence;
\begin{equation}
     \mathcal{L}_{\boldsymbol{y}}({\boldsymbol{n}}) = \int d\theta\, p(\theta|{\boldsymbol{n}}, \boldsymbol{y}) \mathcal{L}[\tilde{\theta}_{\boldsymbol{y}}({\boldsymbol{n}}), \theta] 
     \label{eq:n-loss}
\end{equation}
is the loss associated with the vector of outcomes $\boldsymbol{n}$, based on the posterior $p(\theta|{\boldsymbol{n}}, \boldsymbol{y})$; $\boldsymbol{y}$ is a vector of control parameters; and $\tilde{\theta}_{\boldsymbol{y}}(\boldsymbol{n})$ is an estimator function, whose form we need to decide.
To do so, we minimise Eq.~\eqref{eq:loss-fom} with respect to $\tilde{\theta}_{\boldsymbol{y}}(\boldsymbol{n})$ using variational calculus. 
This identifies the \emph{optimal} estimator $\tilde{\vartheta}_{\boldsymbol{y}}({\boldsymbol{n}})$, in the sense of leading to minimal information loss, as solution to \cite{jaynes2003probability, riley2006mathematical}
\begin{equation}
    \int d\theta\,p(\theta| \boldsymbol{n}, \boldsymbol{y}) 
    \,\partial_z \mathcal{L}(z,\theta)\vert_{z =  \tilde{\vartheta}_{\boldsymbol{y}}(\boldsymbol{n})} = 0.
    \label{eq:opt-est-eq}
\end{equation}

To solve Eq.~\eqref{eq:opt-est-eq}, a specific loss function must be chosen. 
Phase estimation, for instance, deals with circular parameters and so it relies on trigonometric losses such as $\mathcal{L}(\tilde{\theta},\theta) = 4\,\mathrm{sin}^2[(\tilde{\theta} - \theta)/2]$ \cite{demkowicz2011optimal}.
For parameters other than phases, the quadratic loss 
\begin{equation}
    \mathcal{L}_f(\tilde{\theta},\theta) = [f(\tilde{\theta}) - f(\theta)]^2
    \label{eq:quadratic-loss}
\end{equation}
discussed in the main text can account for errors in a wide range of information-theoretic scenarios \cite{tsang2022generalized, tsang2023operational}.
Here, $f$, identified as a symmetry function in this work, maps the hypothesis $\theta$ into a location (or shift) hypothesis $f(\theta)$ \cite{rubio2024first}, in the sense that the aforementioned $\mathcal{L}_f$ is invariant under translations $f(\theta') = f(\theta) + c$, with arbitrary constant $c$ \cite{jaynes2003probability, linden2014bayesian}.
Those parameters for which $f$ exists are thus referred to as location-isomorphic \cite{rubio2024first}.
Use of the loss function \eqref{eq:quadratic-loss} allows the variational problem in Eq.~\eqref{eq:opt-est-eq} to be straightforwardly solved, finding the optimal estimator function $\tilde{\vartheta}_{\boldsymbol{y},f}(\boldsymbol{n})$ defined in Eq.~(3) of the main text.
The optimality of this estimator holds for arbitrary parameter ranges, sample sizes, priors, and likelihood functions.

To calculate the empirical loss incurred by this estimator, $\mathcal{L}_{\boldsymbol{y},f}({\boldsymbol{n}})$, given in Eq.~(5) in the main text, we insert Eqs.~\eqref{eq:quadratic-loss} and $\tilde{\vartheta}_{\boldsymbol{y},f}(\boldsymbol{n})$ into Eq.~\eqref{eq:n-loss}.
By itself, $\mathcal{L}_{\boldsymbol{y},f}({\boldsymbol{n}})$ does not offer a clear visualisation of the quality of our estimates, but it can be used to construct a Bayesian analogue of error bars.
One approach is to note that, when $\tilde{\theta} \approx \theta$, the loss function \eqref{eq:quadratic-loss} can be Taylor-expanded and the square error written as $(\tilde{\theta} - \theta)^2 \approx \mathcal{L}_f(\tilde{\theta},\theta)/f'(\theta)^2$.
Motivated by this, we choose the error bar $\Delta \tilde{\vartheta}_{\boldsymbol{y},f}(\boldsymbol{n})$, given in Eq.~(4) in the main text, as a formal analogue of this idea, and we employ it as a symmetry-informed, dimensionally consistent standard deviation.
Since this construction relies explicitly on the optimal estimator in Eq.~(3) of the main text, $\Delta \tilde{\vartheta}_{\boldsymbol{y},f}(\boldsymbol{n})$ also implicitly depends on the loss function~\eqref{eq:quadratic-loss}.

We further observe that, strictly speaking, error bars are most appropriate in the context of Gaussian-like distributions, but, for more general cases, they serve only as a heuristic and become less reliable with limited data.
Nevertheless, we adopt this representation of Bayesian errors for consistency with current practice.

Finally, to derive the precision gain quantifier $\mathcal{G}_{\boldsymbol{y}, f}$, given in Eq.~(8) in the main text for a single control parameter, we insert Eqs.~\eqref{eq:quadratic-loss} and $\tilde{\vartheta}_{\boldsymbol{y},f}(\boldsymbol{n})$ into the average loss \eqref{eq:loss-fom}. 
This renders the minimum average loss $\bar{\mathcal{L}}_{\boldsymbol{y}, f, \mathrm{min}} = \int d\theta \, p(\theta) f(\theta)^2 - \mathcal{G}_{\boldsymbol{y}, f}$.
Since the first term is fully determined once the prior $p(\theta)$ and the symmetry function $f$ are specified, reducing the average loss amounts to maximise the precision gain $\mathcal{G}_{\boldsymbol{y}, f}$ over the control parameters $\boldsymbol{y}$.

\bigskip

\noindent \textbf{Derivation of multiparameter ignorance priors}
\smallskip

In this section we derive the multiparameter ignorance prior $p_{\mathrm{MI}}(\varphi, \theta)$, given in Eq.~(14) in the main text. 
This derivation provides a rationale that can be applied to other experimental platforms.

We seek the prior $p_{\mathrm{MI}}(\varphi, \theta|\nu)$ representing a meaningful notion of maximum ignorance about the unknown parameters of our platform, which may in principle be conditioned on $\nu$.
To find it, we impose two physically motivated conditions:
\begin{enumerate}[itemsep=0pt]
    \item $p_{\mathrm{MI}}(\varphi', \theta'|\nu)d\varphi' d\theta' = p_{\mathrm{MI}}(\varphi, \theta|\nu)d\varphi d\theta$, which amounts to demand invariance under the two-parameter symmetry in Eq.~(13) in the main text;
    \item the laser frequency $\nu$ does not inform the value of the unknown parameters. 
\end{enumerate}

Condition $1$ implies the functional equation
\begin{equation}
    \gamma\,p_{\mathrm{MI}}\left[\gamma \varphi, \theta + \frac{\log(\gamma)}{\zeta_{\nu}} \Bigg \vert \nu\right] = p_{\mathrm{MI}}(\varphi, \theta | \nu).
\end{equation}
By taking the derivative with respect to $\gamma$ on both sides, we can transform it into the partial differential equation
\begin{equation}
    \zeta_{\nu}\,p_{\mathrm{MI}}(x, y|\nu) 
    + \zeta_{\nu}x\,\partial_xp_{\mathrm{MI}}(x, y|\nu)
    + \partial_y\,p_{\mathrm{MI}}(x, y|\nu)
    = 0.
\end{equation}
Solving for $p_{\mathrm{MI}}(x, y|\nu)$, we find
\begin{equation}
    p_{\mathrm{MI}}(\varphi,\theta|\nu) = 
    \mathrm{e}^{-\zeta_{\nu}\theta}
    h(\varphi\,\mathrm{e}^{-\zeta_{\nu}\theta}),
    \label{eq:prior-multi-supp}
\end{equation}
where $h$ is an arbitrary function. 
As such, and unlike in many single-parameter scenarios \cite{jaynes2003probability, linden2014bayesian}, condition $1$ does not fully determine the ignorance prior we seek. 

Condition $2$, which allows us to drop the dependency of the prior on the laser frequency as $p_{\mathrm{MI}}(\varphi,\theta|\nu) \rightarrow p_{\mathrm{MI}}(\varphi,\theta)$ \cite{jaynes2003probability}, suffices to address this problem.
This condition is already satisfied for $\theta = 0$, since $p_{\mathrm{MI}}(\varphi,0) = h(\varphi)$. 
For $\theta \neq 0$, we can enforce the aforementioned independence by setting the derivative of Eq.~\eqref{eq:prior-multi-supp} with respect to $\nu$ to zero.
Provided that $\nu \neq \nu_{\mathrm{r}}$, so that $d\zeta_{\nu}/d\nu \neq 0$, we find
\begin{equation}
        h(z) + z\,h'(z) = 0,
    \end{equation}
whose solution is $h(z) \propto 1/z$.
Therefore, the prior representing maximum ignorance about the unknown parameters of our cold atom platform is
\begin{equation}
    p_{\mathrm{MI}}(\varphi,\theta) \propto \frac{1}{\varphi}.
    \label{eq:ig-prior-supp}
\end{equation}
The prior used for the adaptive protocol---and, for consistency, for the other two Bayesian methods---is recovered by normalising Eq.~\eqref{eq:ig-prior-supp} over the prior range $[\varphi_{\mathrm{min}},\varphi_{\mathrm{max}}] \times [\theta_{\mathrm{min}},\theta_{\mathrm{max}}]$.

\bigskip

\noindent \textbf{Experimental implementation}
\smallskip

Atoms are loaded into the fibre-atom junction via the following procedure. A cloud of $\sim10^8$ $^{133}$Cs atoms is cooled in a magneto-optical trap and transferred into a red-detuned optical dipole trap formed using 30\,W of light at 1064\,nm, focussed to a waist of 7\,$\mu$m. Approximately 10$^5$ atoms are thus captured in an elongated dipole trap that overlaps with the fibre junction, perpendicular to the fibre core. The number of atoms in this region of overlap is determined by sending a 10\,$\mu$s pulse of light through the fibre, at a frequency that is close to resonance with the atomic D2 line transition, and measuring the fraction of the light that is transmitted using a single-photon counter. 
A detector dark count rate of $n\sim1$ is measured independently and subtracted from both signals. 

\begin{figure}[t]
    \centering
    \includegraphics[width=\linewidth]{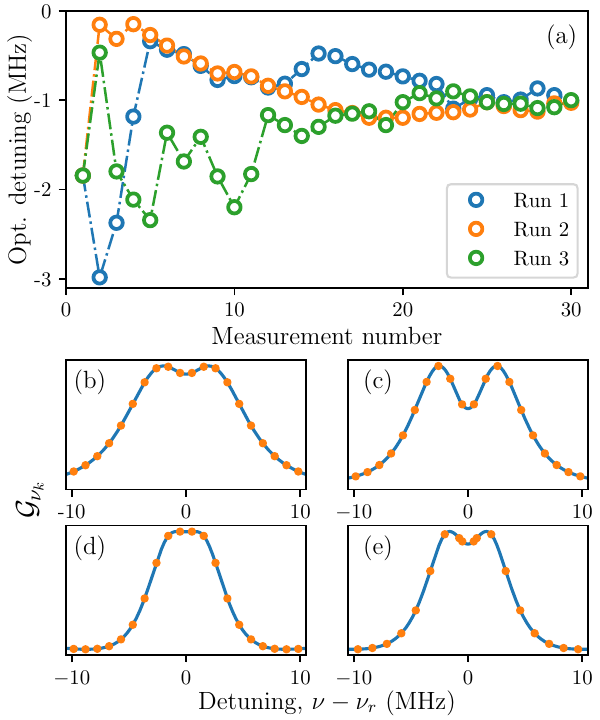}
        \caption{\sf (a) Example of how the optimal measurement frequency $\nu$ changes when the adaptive Bayesian feedback is applied. Three consecutive experimental runs are shown, where the sample has been prepared with an atom number of $N \sim 250$. 
        (b-e) Examples of the gain function $\mathcal{G}_{\nu_k}$ used to find the optimal frequency [Eq.~(8) in the main text]. Examples are shown for: (b) the \emph{a priori} case, $k=1$; (c) high current atom number estimate, $N\sim340$ and $k=4$; (d) low current atom number estimate, $N\sim200$ and $k=4$; (e) at the end of a run, $N\sim250$ and $k=30$. Orange dots show the calculated values, while the blue line is a cubic interpolation.
        }
    \label{fig:frequency-change}
\end{figure}

The small mode area of the intersection ($\sim 20$\,\textmu m$^2$) together with the saturability of the atomic medium limits the applied optical power to $\sim$\,pW, while expansion of the atomic cloud restricts the exposure duration. Total photon counts are thus limited to a few tens of photons per measurement step and photon-counting statistics is the limiting factor on measurement precision.
Typically, averages of $30$ - $100$ measurement shots are recorded, each shot taking approximately $6$\,s. These requirements are comparable to, e.g., Refs.~\cite{Troullinou2023} and \cite{Li_2024,sayrin2015storage}. 
Tab.~\ref{tab:2} below expands the results for all methods in the main text.

\begin{table}[h!]
\centering
\begin{tabular}{l c c c c c }
\hline
\hline
 & \\[-2.75ex]
 &  $m$  & \;$\langle \widetilde{N}_k \rangle$ \; &  $[\mathrm{Var}(\widetilde{N}_k)]^{\frac{1}{2}}$ & NSR & $\langle \Delta \widetilde{N}_k \rangle$ \\[0.25ex]
\hline
On resonance & $16$ & $279$ & $26$ & $0.89\,\%$ & $27$\\
Detuned & $12$ & $313$ & $77$ & $6.10\,\%$ & 47\\
On resonance Bayesian & $16$ & $268$ & $21$ & $0.62\,\%$ & 19\\
\emph{A priori} Bayesian & $19$ & $238$ & $21$ & $0.77\,\%$ & $16$\\
Adaptive Bayesian & $16$ & $258$ & $11$ & $0.19\,\%$ & $18$\\
\hline
\hline
\end{tabular}
    \caption{\sf Overview of parameters of Fig.~2 in the main text, including: number of repeats $m$; mean of the individual estimates, $\langle \widetilde{N}_k \rangle$; the correspondent standard deviation, $[\mathrm{Var}(\widetilde{N}_k)]^{\frac{1}{2}}$; empirical noise-to-signal ratio; and mean of the individual error bars of each estimate $\widetilde{N}_k$ from $m$ repeats, $\langle \Delta \widetilde{N}_k\rangle$.}
     \label{tab:2}
\end{table}
To implement the adaptive Bayesian strategy, the precision gain quantifier $\mathcal{G}_{\nu_k}$ [Eq.~(8) in the main text] needs to be maximised and the optimal frequency calculated in real time and adapted physically sufficiently fast between the measurement shots. Optimal frequency computation was achieved at sufficient speed in Python\,3 on a standard desktop computer. 
Standard techniques, such as caching of function outputs, control of error tolerances and interpolation were employed to reduce evaluation time. In addition, adaptive truncation of the probability distributions for $\theta$, $\bar{n}_b$ and $n$ (the on-resonance optical depth hypothesis, expected photon counts without atoms and photon counts recorded from a specific shot, respectively) was used. 
Truncation ranges were set such that, using the posterior probability distribution in light of all previous measurements, we integrated only over regions where the probability densities were greater than $1\,\%$ of their maximum values, with sums over possible $n$ truncated at the same probability threshold, but under the worst-case assumption for $\theta$ and $\bar{n}_b$, i.e., the sum was only truncated once this threshold was passed for all possible values of $\theta$ and $\bar{n}_b$ remaining within the truncation window. 

The laser frequency was adjusted via a voltage-controlled oscillator within an optical beat lock that stabilises the laser frequency. An example of how the gain function and hence optimal laser frequency changes over the course of a measurement run when the adaptive protocol is applied is shown in Fig.~\ref{fig:frequency-change}. 
Unwanted changes in probe beam power can sometimes result from laser frequency tuning, but in our case these correspond to a fractional power variation of $< 1.5\times10^{-4}$ over the relevant frequency range, which is sufficiently small to be neglected in our analysis. 

\bigskip

\noindent \textbf{Measurement process and example}
\smallskip

\begin{figure}[b]
    \centering
    \includegraphics[width=\linewidth]{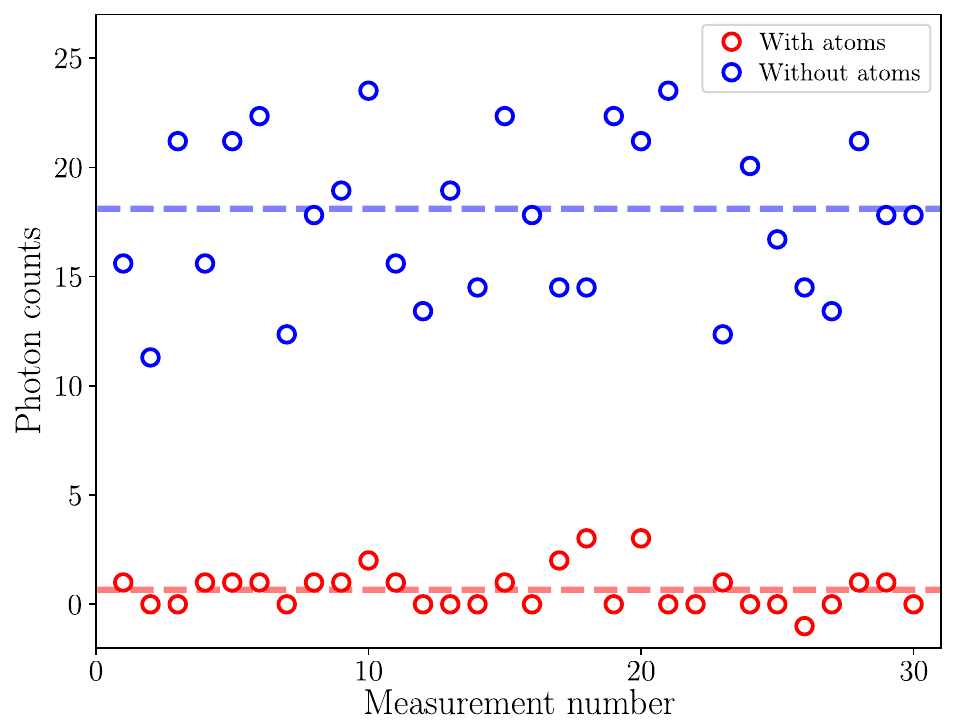}
    \caption{\sf Example of an experimental run consisting of 30 measurements, with atoms present in the hole (red circles) and without (blue circles), using resonant light.}
    \label{fig:data-example}
\end{figure}

Fig.~\ref{fig:data-example} shows an example of the data resulting from the standard measurement process, where the mean number of atoms present in the interaction region was $\sim 250$ and the laser light was applied at the resonant frequency $\nu_\mathrm{r}$. Thirty measurement shots were taken ($k = 30$) with atoms present (red circles), yielding photon counts ${\boldsymbol{n}}_a = (n_{a,1}, \dots, n_{a,k})$, and the same number ($l = 30$) without atoms present (blue circles), yielding photon counts ${\boldsymbol{n}}_b = (n_{b,1}, \dots, n_{b,l})$. Negative photon counts arise due to subtraction of a background dark count rate of $\sim 1$\,count per shot, measured separately, from the data.  

Data taken on-resonance like this can be analysed either via standard maximum likelihood estimation, or via Bayesian analysis. 
For the example in Fig.~\ref{fig:data-example}, $k = 30$ measurements are taken using resonant light with empirical means $\langle n_a \rangle = 0.7 \pm 0.2$ and $\langle n_b \rangle = 18.1 \pm 0.8$. When analysed using the standard protocol, this leads to an atom number estimate of $\tilde{N}_k=283 \pm 22$. When analysed using the Bayesian protocol, as outlined in the main text, it gives an atom number of $\tilde{N}_k=277 \pm 19$. Since these example data were collected with the frequency fixed on resonance, the other protocols do not apply.

\begin{figure}
    \centering
    \includegraphics[width=\linewidth]{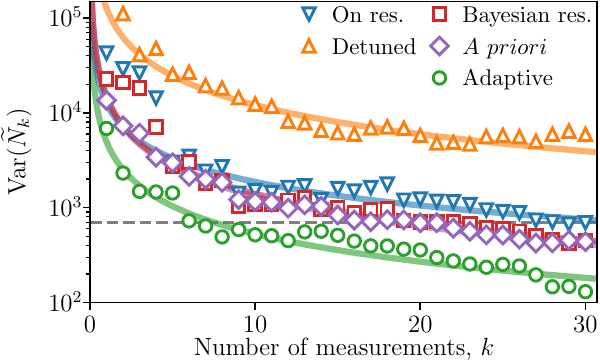}
    \caption{\sf Variance of atom number estimate as a function of number of measurement shots taken for the five methods described in the main manuscript---see main manuscript figure 3 for comparison.}
    \label{fig:variance-variation}
\end{figure}

In Fig.~3 of the main manuscript we presented data showing the noise-to-signal ratio (NSR) as a function of the number of measurement shots taken for the five methods considered. 
Fig.~\ref{fig:variance-variation} details the corresponding results for the variance of the atom number estimate, showing an essentially equivalent trend.

\bigskip

\noindent \textbf{Assessment of expected gain for different experimental settings}

\smallskip

While an adaptive approach is the most robust and resource-effective strategy by construction, it is possible to consider a preliminary estimation of the expected gain before implementation in a new experimental setting.  
A rigorous quantification requires a full determination of the expected information loss as a function of the number of measurements \( k \), integrated over all possible hypotheses \( \theta \) and measurement outcomes \( n \). 
For a quick assessment, local estimation theory provides a practical means to evaluate whether different hypotheses \( \theta \) within the \textit{a priori} distribution \( p_0(\theta) \) necessitate significantly different values of the control parameter \( y \) for optimal measurement. 
Specifically, it allows one to check whether applying the optimal control parameter value(s) associated with one plausible hypothesis \( \theta_a \) would, according to local estimation theory, perform substantially worse under the assumption of another possible hypothesis \( \theta_b \) than using the control parameter value(s) optimal to \( \theta_b \). 

\bigskip

\noindent \textbf{Extensions and next steps}

At a fundamental level, the next theory step is implementing not only optimal estimators, but finding optimal states $\rho_y(\theta)$ and measurements $\mathcal{M}_y(n)$ with statistics $p(n|\theta, y) = \mathrm{Tr}[\mathcal{M}_y(n) \rho_y(\theta)]$. Optimal measurements are given by projecting onto the eigenspace of an operator $\mathcal{S}_{y,f}$ solution to the Lyapunov equation  
\begin{equation}
    \mathcal{S}_{y,f} \rho_{y,f,0} + \rho_{y,f,0} \mathcal{S}_{y,f} = 2\rho_{y,f,1},
    \label{eq:lyapunov}
\end{equation}
with $\rho_{y,f,l} = \int d\theta\,p(\theta)\rho_y(\theta)f(\theta)^l$ \cite{rubio2024first, tsang2022generalized}, while
optimal states can be found by combining this with techniques such as tensor networks and machine learning \cite{schuld2015an, lukas2021neural, dominik2022neural, gebhart2023learning, rinaldi2024parameter, bavaresco2024designing, kurdzialek2024quantum}. 
Furthermore, the derivation of the ignorance prior \eqref{eq:ig-prior-supp} provides a blueprint for extending the symmetry-informed approach in this work to the simultaneous estimation of multiple parameters \cite{demkowicz2020multiparameter}.

\end{document}